\documentclass[12pt]{article}

\usepackage{graphicx}
\usepackage{amssymb}
\usepackage{times}
\usepackage{epsfig}
\usepackage{float}

\newcommand{\remove}[1]{}
\newtheorem{definition}{Definition}

\newtheorem{cor}{Corollary}
\newtheorem{theorem}{Theorem}
\newtheorem{obs}{Observation}
\newtheorem{lemma}{Lemma}

\newcommand{\qed}{\vrule height6pt width4pt\medskip}

\newcommand{\comment}[1]{}
\newcommand{\junk}[1]{}
\newenvironment{proof}{{\em Proof:\ }}{\hfill$\Box$\medskip}

\begin{document}
\title{Suffix Trays and Suffix Trists: Structures for Faster Text Indexing\footnote{Results from this paper have appeared as an extended abstract in ICALP 2006.}}
\author{
\begin{tabular}{ccc}
Richard Cole\thanks{Department of Computer Science, Courant
Institute, NYU.}\thanks{This work was supported in part by NSF grant CCF-1217989} & Tsvi Kopelowitz\thanks{Dept. of Computer
Science, Bar-Ilan U., 52900 Ramat-Gan, Israel} & Moshe Lewenstein\thanks{This research was supported by a BSF grant (\#2010437) and a GIF grant (\#1147/2011)}
\end{tabular}
}
\date{}

\maketitle

\begin{abstract}
Suffix trees and suffix arrays are two of the most widely used data
structures for text indexing. Each uses linear space and can be
constructed in linear time for polynomially sized alphabets. However,
when it comes to answering queries with worst-case deterministic time bounds, the prior does so in
$O(m\log|\Sigma|)$ time, where $m$ is the query size, $|\Sigma|$ is
the alphabet size, and the latter does so in $O(m+\log n)$ time, where
$n$ is the text size. If one wants to output all appearances of the
query, an additive cost of $O(occ)$ time is sufficient, where $occ$ is the
size of the output. Notice that it is possible to obtain a worst case, deterministic query time of $O(m)$ but at the cost of super-linear construction time or space usage.

We propose a novel way of combining the two into, what we call, a
{\em suffix tray}. The space and construction time remain linear and
the query time improves to $O(m+\log|\Sigma|)$ for integer alphabets from a linear range, i.e. $\Sigma \subset \{1,\cdots, cn\}$, for an arbitrary constant $c$. The construction and query are deterministic. Here also an
additive $O(occ)$ time is sufficient if one desires to output all
appearances of the query.

We also consider the online version of indexing, where the text arrives online, one character at a time, and indexing queries are
answered in tandem. In this variant we create a cross between a
suffix tree and a suffix list (a dynamic variant of suffix array) to
be called a {\em suffix trist}; it supports queries in
$O(m+\log|\Sigma|)$ time. The suffix trist also uses linear space. Furthermore, if there exists an
online construction for a linear-space suffix tree such that the
cost of adding a character is worst-case deterministic $f(n,|\Sigma|)$ ($n$ is the size
of the current text), then one can further update the suffix trist in $O(f(n,|\Sigma|)+\log |\Sigma|)$ time. The best currently known worst-case deterministic bound for $f(n,|\Sigma|)$ is $O(\log n)$ time.
\end{abstract}

\section{ Introduction}\label{sec:intro}

{\em Indexing} is one of the most important paradigms in
searching. The idea is to preprocess a text and construct a
mechanism that will later provide answers to queries of the form
"does a pattern $P$ occur in the text" in time proportional to the
size of the {\sl pattern} rather than the text. The suffix
tree~\cite{F-97,Mc-76,U-95,W-73} and suffix
array~\cite{KS-03,KSPP:03,KA:03,MM-90} have proven to be
invaluable data structures for indexing.

Both suffix trees and suffix arrays use $O(n)$ space, where $n$ is
the text length. In fact, for alphabets from a polynomially sized
range, both can be constructed in linear time,
see~\cite{F-97,KS-03,KSPP:03,KA:03}.

The query time is slightly different in the two data structures.
Namely, using suffix trees queries are answered in
$O(m\log|\Sigma|+occ)$ time, where $m$ is the length of the query,
$\Sigma$ is the alphabet, $|\Sigma|$ is the alphabet size and $occ$
is the number of occurrences of the query. Using suffix arrays the time
is $O(m + \log n + occ)$. In~\cite{CL-03} it was shown that the search time of $O(m + \log n + occ)$ is possible also on suffix trees.
For the rest of this paper we assume that
we are only interested in one occurrence of the pattern in the text,
and note that we can find all of the occurrences of the pattern with
another additive $occ$ cost in the query time.

The differences in the running times follows from the different
ways queries are answered. In a suffix tree, queries are answered by
traversing the tree from the root. At each node one needs to know
how to continue the traversal and one needs to decide between at most
$|\Sigma|$ options which are sorted, which explains the
$O(\log|\Sigma|)$ factor. In suffix arrays one performs a binary
search on all suffixes (hence the $\log n$ factor) and uses
longest common prefix (LCP) queries to quickly decide whether the
pattern needs to be compared to a specific suffix
(see~\cite{MM-90} for full details).

It is straightforward to construct a data structure that will yield optimal
$O(m)$ time to answer queries. This can be done by putting
a $|\Sigma|$ length array at every node of the suffix tree. Hence,
when traversing the suffix tree with the query we will spend
constant time at each node. However, the size of this structure is
$O(n|\Sigma|)$. Also, notice that this method assumes $\Sigma$ is comprised of $\{1,2,...,|\Sigma|-1\}\cup \{\$\}$\footnote{Note that the special $\$ $ character is a special delimiter which has only appears at the end of the text and is considered to be lexicographically larger than all of the other integers in $\Sigma$.} as we need to be able to random access locations in the array based on the current character, and so we need our alphabet to be proper indices. While this can be overcome using renaming schemes, it will provide an extra $O(m\log |\Sigma|)$ time for the query process, as one would need to rename each of the characters in the pattern.

Another variant of suffix trees that will yield optimal
$O(m)$ time to answer queries is as follows. Construct a suffix tree which maintains a static dictionary
at each node, $h_v:\Sigma_v \rightarrow E_v$ where $E_v$ is the set of edges exiting $v$ and $\Sigma_v$ is the set of  first characters on $E_v$ edges. We set $h_v(\sigma) = x$, where $x$ is the edge associated with the first character $\sigma$. This scheme can be built in linear randomized time and linear space. If one desires determinism, as we do, the downside is that the construction of the deterministic static dictionary takes, in the best (and complicated) case, $O(n\min\{\log |\Sigma|,\log\log n\})$ preprocessing time~\cite{ruzic}.

The question of interest here is whether one can deterministically construct an indexing data structure using $O(n)$
space and time that will answer queries faster than suffix arrays and suffix trees. We indeed propose to
do so with the {\em Suffix Tray}, a new data structure that extracts
the advantages of suffix trees and suffix arrays by combining their
structures. This yields an $O(m+\log|\Sigma|)$ query time. However, our solution uses some $|\Sigma|$ length arrays to allow for quick navigation, and as such it seems that we are also confined to using integer alphabets. However, this can be improved to alphabets $\Sigma \subset \{1,\cdots, cn\}$, for arbitrary constant $c$, since the text can be sorted and a renaming array $A$ of length $cn$ can be maintained saving the rank (relative to $\Sigma$) of the element, i.e. $A[\sigma]$ is the rank of $\sigma$ in $\Sigma$ and $A[i]$ is 0 if there is no character $i$ in $T$. For each pattern character evaluated a constant time lookup in $A$ gives the rank of the character.

We also consider the natural extension to the online case, that is the scenario where online update of the
text are allowed. In other words, given an indexing structure supporting
indexing queries on $S$, we would also like to support extensions
of the text to $Sa$, where $a\in\Sigma$. We assume that the text
is given in reverse, i.e. from the last character towards the
beginning. So, an indexing structure of our desire when
representing $S$ will support extensions to $aS$ where $a \in
\Sigma$. We call the change of $S$ to $aS$ a {\em text extension}.
The "reverse" assumption that we use is not strict, as most
indexing structures can handle online texts that are reversed
(e.g. instead of a suffix tree one can construct a prefix tree and
answer the queries in reverse. Likewise, a prefix array can be
constructed instead of a suffix array).

Online constructions of indexing structures have been suggested
previously. McCreight's suffix tree algorithm~\cite{Mc-76} was the
first online construction. It was a reverse construction (in the
sense mentioned above). Ukkonen's algorithm~\cite{U-95} was the
first online algorithm that was not reversed. In both these
algorithms text extensions take $O(\log |\Sigma|)$ amortized time, but in the worst-case a text extension could take $\Omega(n\log|\Sigma|)$ time. In~\cite{AKLL-05} an online suffix tree
construction (under the reverse assumption) was proposed with
$O(\log n)$ worst-case time per text extension. In all of these constructions
a full suffix tree is constructed and hence queries are answered
in $O(m\log|\Sigma|)$ time. An online variant of suffix arrays
was also proposed in~\cite{AKLL-05} with $O(\log n)$ worst-case
time per text extension and $O(m + \log n)$ time for answering queries.
Similar results can be obtained by using the results
in~\cite{gi-99}.

The problem we deal with in the second part of the paper is how to
build an indexing structure that supports both text extensions (to the beginning of the text) and
supports fast(er) indexing. We will show that if there exists an
online construction for a linear-space suffix tree such that the
cost of adding a character is $f(n,|\Sigma|)$ ($n$ is the size
of the current text), then we can construct an online linear-space
data-structure for indexing that supports indexing queries in
$O(m+\log |\Sigma|)$ time, where the cost of adding a character is
$O(f(n,|\Sigma|)+\log |\Sigma|)$ time. We will call this data structure
the {\em Suffix Trist}.

As mentioned in the previous paragraph the best currently known worst-case deterministic bound for $f(n,|\Sigma|)$ is $O(\log n)$ time~\cite{AKLL-05}.
Breslauer and Italiano~\cite{BI:13} have improved this result for the case that $|\Sigma|$ is $o(\log n)$. Specifically,
they show how to support text extensions in deterministic $O(|\Sigma| + \log\log n)$ time. In~\cite{Kop:12} an indexing data structure was shown
for which text extensions can be implemented in $O(\log\log n +\log\log |\Sigma|)$ expected time. However, the construction there is randomized and
does not suit our needs.

\section{Suffix Trees, Suffix Arrays and Suffix Intervals}\label{sec:suffix_intervals}

Consider a text $S$ of length $n$ and let $S^1,\cdots, S^n$ be the
suffixes of $S$. Two classical data structures for indexing are
the suffix tree and the suffix array. It is assumed that the reader is
familiar with the suffix tree. Let $S^{i_1},...,S^{i_n}$ be the
lexicographic ordering of the suffixes. The suffix array of $S$ is
defined to be $SA(S)= <i_1,...,i_n>$, i.e. the indices of the
lexicographic ordering of the suffixes.
Location $j$ of the suffix array is sometimes referred to as the location of $S^{i_j}$
(instead of the location of $i_j$).

\begin{figure}
\begin{center}
  \includegraphics[width=4.25in]{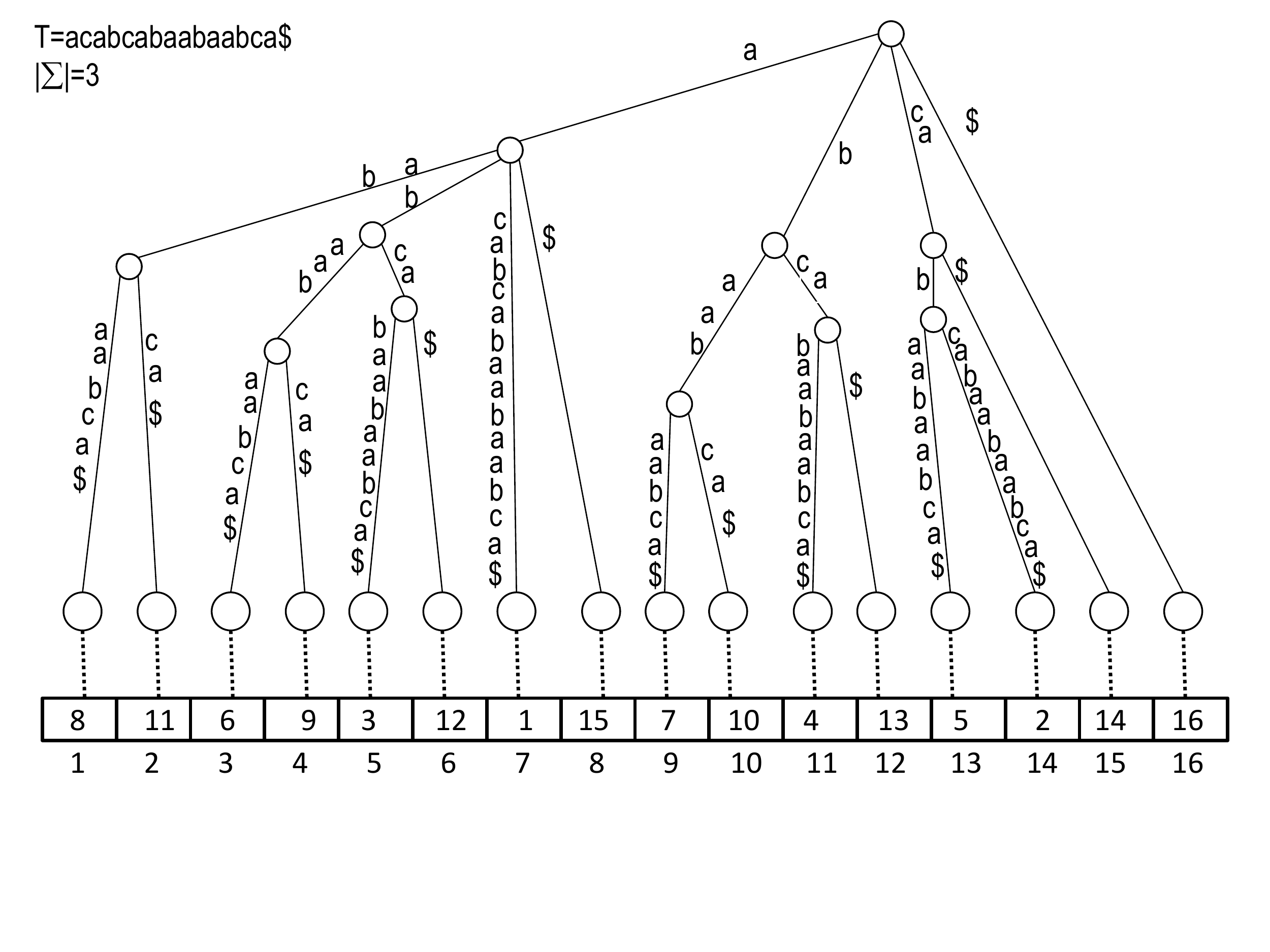}
  \caption{\sf\small A suffix tree with a suffix array at leaves.}\label{fig1}
\end{center}
\end{figure}

Let $ST(S)$ and $SA(S)$ denote the suffix tree and suffix array
of $S$, respectively. As with all suffix tree constructions to
date, assume that every node in a suffix tree
maintains its children in lexicographic order. Therefore, the leaves
ordered by an inorder traversal correspond to the suffixes in
lexicographic order, which is also the order maintained in the
suffix array. Hence, one can view the suffix tree as a tree over the
suffix array. See Figure~\ref{fig1}.

We further comment on the connection between suffix arrays and suffix
trees. For strings $R$ and $R'$ say that $R<_LR'$ if $R$ is
lexicographically smaller than $R'$.  $leaf(S^i)$ denotes the leaf
corresponding to $S^i$ in $ST(S)$, the suffix tree of $S$. Define
$L(v)$ to be the location  of $S^i$ in the suffix array where $leaf(S^i)$ is the leftmost leaf of the subtree
of $v$.
Notice that since the children of a node in a suffix
tree are assumed to be maintained in lexicographic order, it follows that for all
$S^j$ such that $leaf(S^j)$ is a descendant of $v$, $S^i \leq_L
S^j$. Likewise, define $R(v)$  to be the location  of $S^i$ in the suffix array where  $leaf(S^i)$ is the
rightmost leaf of the subtree of $v$. Therefore, for all $S^j$ such
that $leaf(S^j)$ is a descendant of $v$, $S^i \geq_L S^j$. Hence,
the interval $[L(v),R(v)]$ is an interval of the suffix array which
contains exactly all the suffixes $S^j$ for which $leaf(S^j)$ is a
descendant of $v$. Notice that interval representations have already been discussed in~\cite{AKO-04}.

Moreover, under the assumption that the children of a node in a
suffix tree are maintained in lexicographic ordering one can state
the following.

\begin{obs}\label{children-interval}
Let $S$ be a string and $ST(S)$ its suffix tree. Let $v$ be a node
in $ST(S)$ and let $v_1,...,v_r$ be its children in lexicographic order. Let $1 \leq i \leq
j \leq r$,  and let $[L(v_i),R(v_j)]$ be an interval of the suffix
array. Then $k \in [L(v_i),R(v_j)]$ if and only if $leaf(S^k)$ is in
one of the subtrees rooted at $v_i,...,v_j$.
\end{obs}

This leads to the following concept.

\begin{definition}
Let $S$ be a string and $\{S^{i_1},...,S^{i_n}\}$ be the
lexicographic ordering of its suffixes. The interval $[j,k] =
\{{i_j},...,{i_k}\}$, for $j\leq k$, is called a {\em suffix
interval}.
\end{definition}

Obviously, suffix intervals are intervals of the suffix array. Notice that, as mentioned above, for a node $v$ in a suffix tree,
$[L(v),R(v)]$ is a suffix interval and the interval is called {\em
$v$'s suffix interval}. Also, by Observation~\ref{children-interval}, for
$v$'s children $v_1,...,v_r$ and for any $1 \leq i \leq j \leq r$,
we have that $[L(v_i),R(v_j)]$ is a {\em suffix interval} and this
interval is called the {\em $(i,j)$-suffix interval}.

\section{Suffix Trays}\label{sec:suffix_trays}

The suffix tray is now introduced. The suffix tray will use the
concept of suffix intervals from the previous section which, as has
been seen, forms a connection between the nodes of the suffix trees and
intervals in the suffix array.

For suffix trays, special nodes are created, which correspond to
suffix intervals. These nodes are called {\em suffix interval nodes}.
Notice that these nodes are not suffix tree nodes.
Part of the suffix tray will be a suffix array. Each suffix interval
node can be viewed as a node that maintains the endpoints of the
interval within the complete suffix array.

Secondly, the idea of the space-inefficient $O(n|\Sigma|)$
suffix tree solution mentioned in the introduction is used. We construct
$|\Sigma|$-length arrays for a selected subset of nodes, a subset
that contains no more than ${n\over|\Sigma|}$ nodes, which retains
the $O(n)$ space bound. To choose this selected subset of nodes we define the following.

\begin{definition} Let $S$ be a string over alphabet $\Sigma$.
A node $u$ in $ST(S)$ is called a $\sigma$-node if the number of
leaves in the subtree of $ST(S)$ rooted at $u$ is at least
$|\Sigma|$.\ \  A $\sigma$-node $u$ is called a
branching-$\sigma$-node, if at least two
 of $u$'s children in $ST(S)$ are $\sigma$-nodes and is called a $\sigma$-leaf if all
 its children in $ST(S)$ are not $\sigma$-nodes.
\end{definition}

\begin{figure}[H]
\begin{center}
  \includegraphics[width=4.25in]{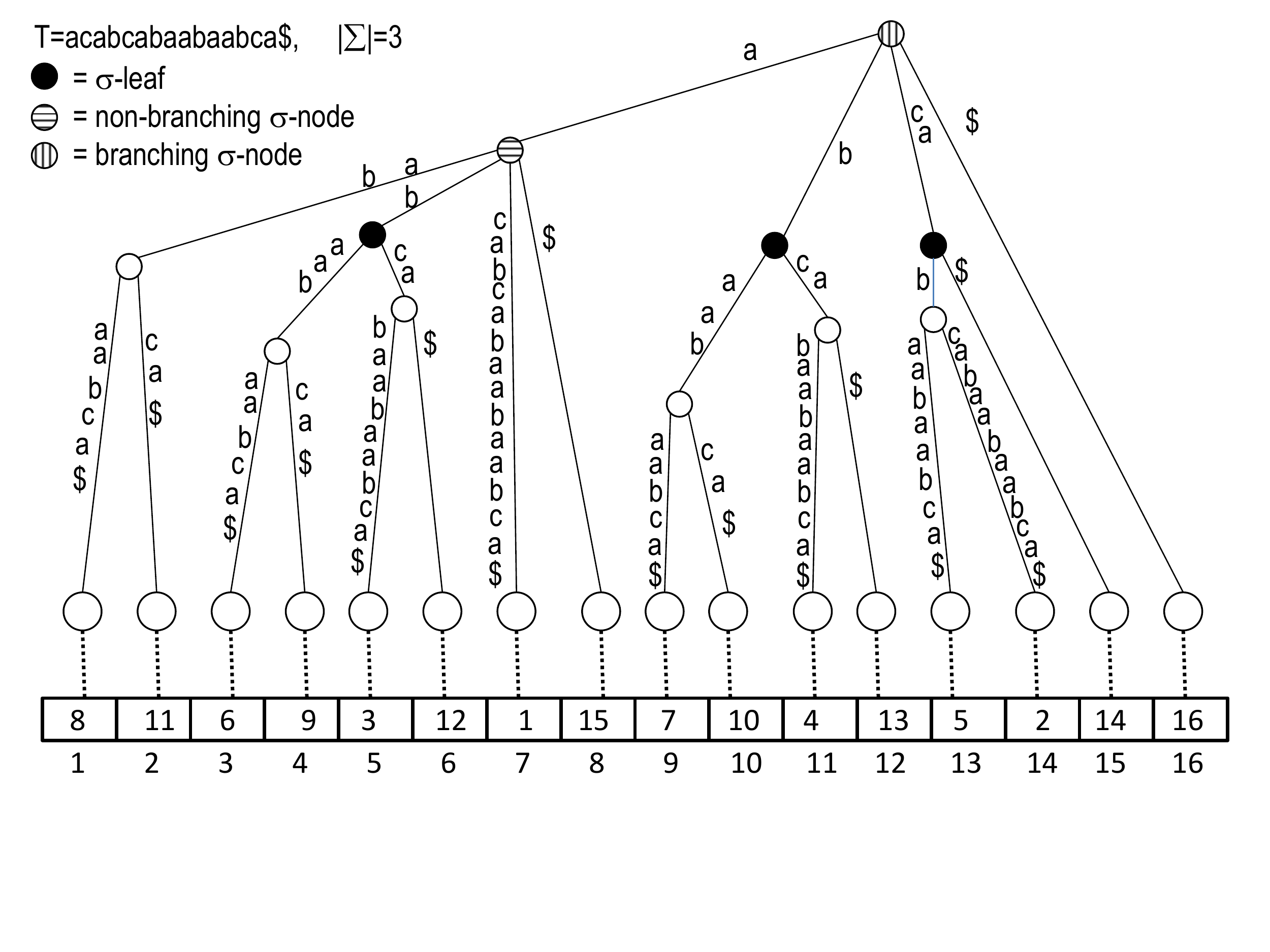}
  \caption{\sf A suffix tree with different node types.}\label{fig2}
\end{center}
\end{figure}

See Figure~\ref{fig2} for an illustration of the different node types.

The following property of branching-$\sigma$-nodes is crucial to our result.

\begin{lemma}\label{branching-nodes-size}
Let $S$ be a string of size $n$ over an alphabet $\Sigma$ and let
$ST(S)$ be its suffix tree. The number of branching-$\sigma$-nodes
in $ST(S)$ is $O({n \over |\Sigma|})$.
\end{lemma}

\proof The number of $\sigma$-leaves is at most ${n \over |\Sigma|}$
because (1) they each have at least $|\Sigma|$ leaves in their
subtree and (2) their subtrees are disjoint. Let $T$ be the tree
induced by the $\sigma$-nodes and contracted onto the
branching-$\sigma$-nodes and $\sigma$-leaves only. Then $T$ is a
tree with ${n \over |\Sigma|}$ leaves and with every internal node
having at least 2 children. Hence, the lemma follows. \qed

This implies that one can afford to construct arrays at every
branching-$\sigma$-node which will be used for answering
queries quickly as shall be seen in Section~\ref{sub_sec:query_navigation}.

\junk{

\begin{definition} For a string $S$
over alphabet $\Sigma$, the $\sigma$-tree is the subtree of $ST(S)$
induced by the $\sigma$-nodes of $ST(S)$. Denote the
$\sigma$-tree by $\sigma$-$ST(S)$.
\end{definition}
}

\subsection{Suffix Tray Construction}\label{sub_sec:suffix_tray_construction}

A {\em suffix tray} is constructed from a suffix tree as follows.
The suffix tray contains all the $\sigma$-nodes of the suffix tree.
Some suffix interval nodes are also added to the suffix tray as
children of $\sigma$-nodes. Here is how each $\sigma$-node is
converted from the suffix tree to the suffix tray.

\begin{itemize}
\item \underline{$\sigma$-leaf $u$:} $u$ becomes a suffix
interval node with suffix interval $[L(u),R(u)]$.
\item \underline{non-leaf $\sigma$-node $u$:} Let $u_1,...,u_r$ be
$u$'s children in the suffix tree and $u_{l_1},...,u_{l_x}$ be the
subset of children that are $\sigma$-nodes. Then $u$ will be in the
suffix tray and its children will be interleaving suffix interval nodes and
$\sigma$-nodes, i.e. $(1,l_1-1)$-suffix interval node, $u_{l_1}$,
$(l_1+1,l_2-1)$-suffix interval node, $u_{l_2}$, ..., $u_{l_x}$,
$(l_x+1,r)$-suffix interval node.
\end{itemize}

At each branching-$\sigma$-node $u$ in the suffix tray we construct
an array of size $|\Sigma|$, denoted by $A_u$, that contains the
following data. For every child $v$ of $u$ that is a
$\sigma$-node, location $\tau$ in $A_u$ where $\tau$ is the first
character on the edge $(u,v)$, points to $v$. The rest of the
locations in $A_u$ point to the appropriate suffix interval node,
or to a NIL pointer if no such suffix interval exists.

\begin{figure}
\begin{center}
  \includegraphics[width=4.25in]{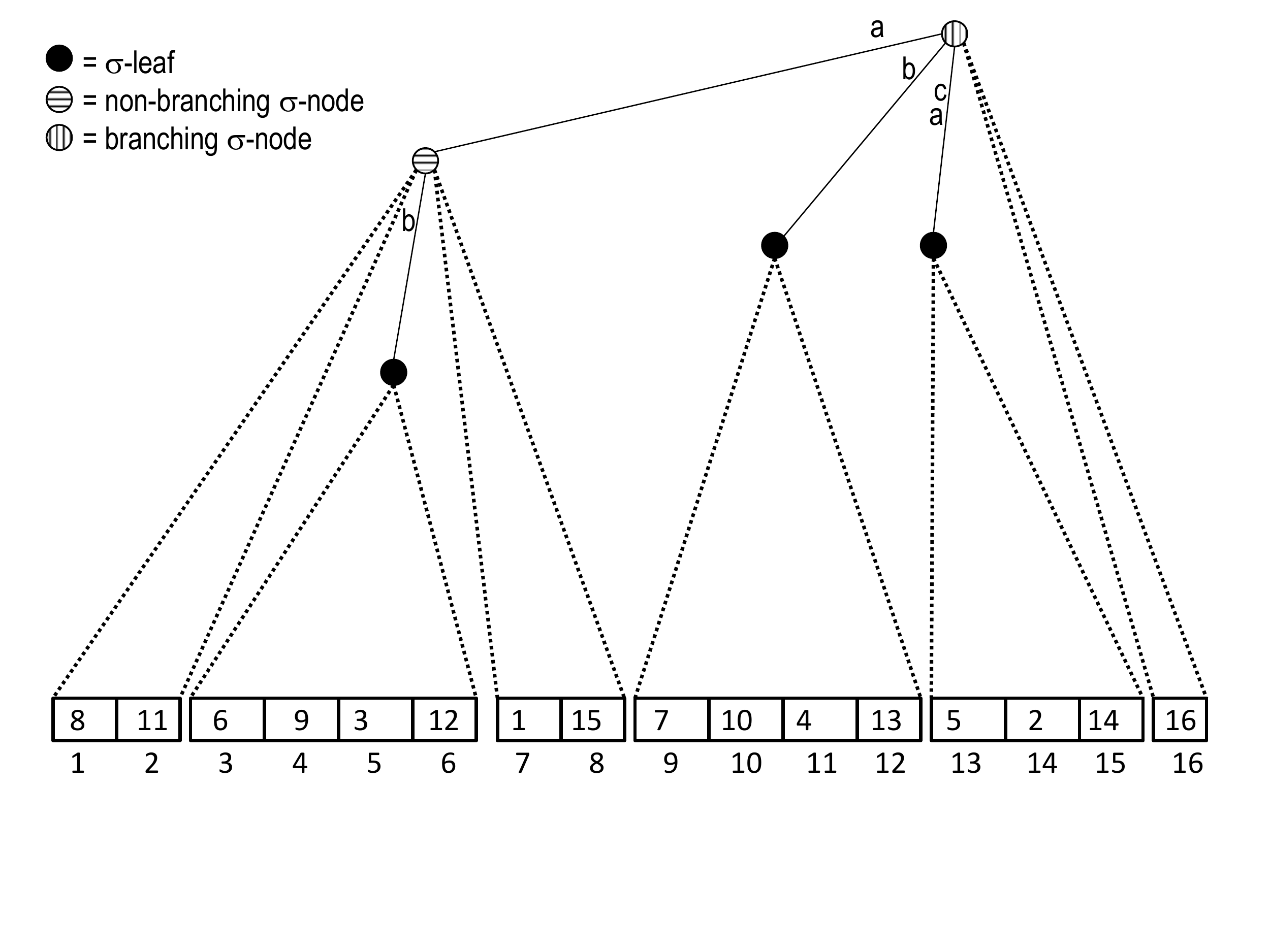}
 \caption{\sf \small A suffix tray (on running example) with chunks of suffix array at the bottom.}\label{fig3}
\end{center}
\end{figure}

At each $\sigma$-node $u$ which is not a branching-$\sigma$-node
and not a $\sigma$-leaf, i.e. it has exactly one child $v$ which
is a $\sigma$-node, store the first character $\tau$ on the
edge $(u,v)$, which is called the {\em separating character}, together with two pointers to its two interleaving suffix intervals.
See Figure~\ref{fig3} for an example of a suffix tray.

The suffix tray is now claimed to be of linear size.

\begin{lemma}
Let $S$ be a string of size $n$. Then the size of the suffix tray
for $S$ is $O(n)$.
\end{lemma}

\proof The suffix array is clearly of size $O(n)$ and the number of
suffix interval nodes is bounded by the number of nodes in $ST(S)$.
Also, for each non branching-$\sigma$-node the auxiliary information
is of constant size.

The auxiliary information held in each branching-$\sigma$-node is of
size $O(|\Sigma|)$. By Lemma~\ref{branching-nodes-size} there are
$O({n \over |\Sigma|})$ branching-$\sigma$-nodes. Hence, the overall size is $O(n)$. \qed

Obviously, given a suffix tree and suffix array, a suffix tray can
be constructed in linear time (using depth-first searches, and
standard techniques). Since both suffix arrays and suffix trees
can be constructed in linear time for alphabets from a
polynomially sized range~\cite{F-97,KS-03,KSPP:03,KA:03}, so can
suffix trays.

\subsection{Navigating on Index Queries}\label{sub_sec:query_navigation}

Suffix trays main goal is answering index queries. Now we explain how to do so.

Upon receiving a query $P=p_1...p_m$, we traverse the
suffix tray from the root. Say that we have already traversed the suffix tray with $p_1...p_{i-1}$ and need to continue with
$p_i...p_m$. At each branching-$\sigma$-node $u$, access array $A_u$ at
location $p_i$ in order to determine which suffix
tray node to navigate to. Obviously, since this is an array lookup
this takes constant time. For other $\sigma$-nodes that are not
$\sigma$-leaves and not branching-$\sigma$-nodes, compare $p_i$
with the separator character $\tau$. Recall that these nodes have
only one child $v$ that is a $\sigma$-node. Hence, the children of $u$ in the suffix
tray are (1) a suffix interval node to the
left of $v$, say $u$'s left interval, (2) $v$, and (3) a suffix
interval node to the right of $v$, say $u$'s right interval. If
$p_i < \tau$, navigate to $u$'s left interval. If $p_i
> \tau$, navigate to $u$'s right interval. If $p_i = \tau$, navigate to the only child of $u$ that is a $\sigma$-node. If $u$ is
a $\sigma$-leaf then we search within $u$'s suffix interval.

To search within a suffix interval $[j,k]$, the
suffix array search is applied beginning with boundaries $[j,k]$\footnote{This can easily be done with the RMQ data structure. However, the original result of Manber and Myers~\cite{MM-90} is slightly more rigid. The more expansive view is described in~\cite{Lew-13}.}. The time to
search in this structure is $O(m+\log I)$, where $I$ is the interval
size. Hence, the following lemma is helpful.

\begin{lemma}\label{bound-intervals}
Every suffix interval in a suffix tray is of size $O(|\Sigma|^2)$.
\end{lemma}

\proof Consider an $(i,j)$-suffix interval, i.e. the interval
$[L(v_i),R(v_j)]$ which stems from a node $v$ with children
$v_1,...,v_r$. Notice that, by Observation~\ref{children-interval}, the
$(i,j)$-suffix interval contains the suffixes which are represented
by leaves in the subtrees of $v_i,...,v_j$. However, $v_i,...,v_j$
are not $\sigma$-nodes (by suffix tray construction). Hence, each
subtree of those nodes contains at most $|\Sigma|-1$ leaves. Since
$j-i+1\leq |\Sigma|$ the overall size of the $(i,j)$-suffix interval
is $O(|\Sigma|^2)$.

A suffix interval $[L(v),R(v)]$ is maintained only for
$\sigma$-leaves. As none of the children of $v$ are  $\sigma$-nodes
this is a special case of the $(i,j)$-suffix interval. \qed

By the discussion above and Lemma~\ref{bound-intervals} the running
time for answering an indexing query is $O(m+\log|\Sigma|)$.
In summary we have proven the following.

\begin{theorem} Let $S$ be a length $n$ string over an alphabet $\Sigma \subset \{1,\cdots, cn\}$.
The suffix tray of $S$ is (1) of size $O(n)$, (2) can be constructed
in $O(n)$ time, and (3) supports indexing queries (of size $m$) in $O(m+\log|\Sigma|)$ time.
\end{theorem}

\section{Suffix Trists - The Online Scenario}\label{sec:suffix_trists}

In this section the problem of how to build an indexing
structure that supports both text extensions and supports fast(er)
indexing is addressed. It is shown that if there exists an online construction for a
linear-space suffix tree such that the cost of adding a character is
$f(n,|\Sigma|)$ ($n$ is the size of the current text), then one
can construct an online linear-space data-structure for indexing
that supports indexing queries in $O(m+\log |\Sigma|)$ time, where
the cost of adding a character is $O(f(n,|\Sigma|)+\log |\Sigma|)$ time.
As text extensions occur, the online linear-space
suffix-tree construction is treated as a {\em suffix-tree oracle}; it performs the appropriate updates to the suffix tree as a result of the text extension, and maintains the on-line suffix-tree as a result. Specifically, the
best known current construction supports text extensions in $O(\log
n)$ time (see~\cite{AKLL-05}). When $|\Sigma|$ is small one can obtain an improved $O(|\Sigma| + \log\log n)$ time~\cite{BI:13}.

We propose an online version of the Suffix Tray data structure which we call a {\em Suffix Trist} (a cross between a suffix tree and an enhanced linked list). The suffix trist intends to imitate the suffix tray. The
$\sigma$-nodes and branching-$\sigma$-nodes are still used in the suffix tree part of the suffix trist,
and the method for answering indexing queries is similar. However, new issues arise in the online model:

\begin{enumerate}
\item{Suffix arrays are static data structures and, hence, do not support insertions of new suffixes.}
\item{The arrays $A_v$, for the branching-$\sigma$-nodes, need to be initialized and made dynamic.}
\item{The status of nodes changes as time progresses (non-$\sigma$-nodes become $\sigma$-nodes, and $\sigma$-nodes become branching-$\sigma$-nodes).}
\end{enumerate}

We will show in Section~\ref{sub_sec:BIS} how to overcome (1) by using an alternative data structure. We will have many of these alternative data structures, each representing a suffix interval. In Section~\ref{sub_sec:BIS_insertion} we show how to find the correct alternative data structure (among those representing the different suffix intervals) when implementing text extensions. In Section~\ref{sub_sec:dynamic_branching_array} we propose an alternative way to handling the array $A_v$ of branching-$\sigma$-nodes. Section~\ref{sec:status_change} is dedicated to solving (3) and is a direct worst-case solution.

\subsection{Balanced Indexing Structures for Suffix Trists}\label{sub_sec:BIS}

Since we mimic the Suffix Tray with the Suffix Trist in an online (dynamic) manner we need to show how to replace the (static) suffix arrays with dynamic data structures. On the other hand, we also want to make sure that the new data structures can answer queries efficiently in the worst case.

 The two data structures to be used are: (1) a {\em suffix list} which is a doubly linked list of the lexicographically ordered suffix collection, and (2) a {\em balanced indexing structure, BIS} for short, which is a balanced binary search tree over the lexicographically ordered suffix collection, fully described by Amir et al.~\cite{AKLL-05}. The BIS augments the suffix list with the dynamic order maintenance data structure of Dietz and Sleator~\cite{DS:87} (see~\cite{Kop:12} for a simplification). The BIS contains a suffix list within. Moreover, it contains the longest common prefix (LCP) data structure from~\cite{FG:04} in order to allow indexing queries to be answered efficiently.  A BIS supports a {\em text extension} in $O(\log n)$ time and indexing queries in $O(m+\log n)$ time (see~\cite{AKLL-05} and the next paragraph for a synopsis). The $\log n$ term in the indexing query time follows from the height of the tree.

In a nutshell, a BIS is defined for a string $S$ and supports text extensions to $aS$ and indexing queries. We give an outline of text extensions (queries are similar) from~\cite{AKLL-05}. We "insert" the new suffix $aS$ into the BIS by following the binary search tree from the root to the appropriate location of $aS$ in the tree, as one would for a standard binary search. However, a straightforward comparison may take $\Omega(|S|)$ time as $aS$ and the string represented by a node may be of length $\Omega(|S|)$. To reduce the time to $O(1)$ an LCP (longest common prefix) data structure is used on all suffixes. So, at a node representing suffix $S'$ we compare $aS$ to $S'$ by first comparing their first character. If they are not equal then the walk down the binary search tree continues according to the comparison. If they are equal we use the LCP data structure to compare $S$ and $S''$, where $S'=aS''$. Notice that we compare $S$ and $S''$ and not $aS$ and $S'$ because $aS$ is new and not yet in the LCP data structure. Once $aS$ is in place in the binary search tree one adds it to the LCP data structure. This is done using a method very similar to the results in~\cite{FG:04} and is out of scope of this paper. For more details see~\cite{AKLL-05}.

The above suggests replacing a suffix array of the Suffix Tray with a BIS in the {\em Suffix Trist} thereby creating a separate BIS for every suffix
interval. Since the suffix intervals are of size $O(|\Sigma|^2)$
the search time in the small BISs will be $O(m+\log |\Sigma|)$.

However, while the search is within the desired time bounds, one has to be more careful during the execution of a text extension. There are two issues that need to be considered. The first is that we assume that we are given an online suffix tree construction. Hence, the suffix tree is updated with a new node representing the new suffix of $aS$. However, one needs to find the appropriate BIS into which the new node needs to be inserted. This is discussed in Section~\ref{sub_sec:BIS_insertion}.

The second issue is that each BIS contains only some of the suffixes, whereas the search of the location of $aS$ in the BIS relies on the underlying LCP data structure, described above, which requires that all the suffixes of $S$ be in the BIS (which is almost always not the case). However, this may be solved by letting the LCP data structure work on all suffixes in all BISs. To make this work we need to show how to insert the new suffix $aS$ (the text extension) into the data structure. We observe that once we find the location of $aS$ in the current BIS this will be its location (not only in the current BIS but also) among all suffixes in all BISs. This is true because the BISs represent suffix intervals (i.e. contiguous suffixes).

\subsection{Inserting New Nodes into BISs}\label{sub_sec:BIS_insertion}

When a text extension from $S$ to $aS$ is performed, the suffix tree
is updated to represent the new text $aS$ (by our suffix-tree
oracle). Specifically, a new leaf, corresponding to the new suffix,
is added to the suffix tree, and perhaps one internal node is also
added. If such an internal node is inserted, then that node is the
parent of the new leaf and this happens in the event that the new
suffix diverges from an edge (in the suffix tree of $S$) at a
location where no node previously existed. In this case an edge
needs to be broken into two and the internal node is added at that
point.

We now show how to update the suffix trist using the output of the online suffix tree oracle.
The steps are (1) finding the correct BIS
into which the new suffix is to be inserted and (2) performing the insertion of the new suffix into
this BIS. Of course, this may change the status of internal nodes,
which are handled in Section~\ref{sec:status_change}. The focus is on solving
(1) while mentioning that (2) can be solved by the text extension method of the BIS whose synopsis was given in Section~\ref{sub_sec:BIS}.

The following useful observation is immediately derived from the definition of suffix trists.

\begin{obs}\label{leaf}
For a node $u$ in a suffix tree, if $u$ is not a $\sigma$-node, then
all of the leaves in $u$'s subtree are in the same BIS.
\end{obs}

For every node $u$ in the suffix tree it will be useful to maintain a pointer $leaf(u)$ to some leaf
in $u$'s subtree. This invariant can easily be maintained under text extensions since an internal node is
always created together with a new leaf, and this leaf will always be a leaf of the subtree of the internal node.

In order to find the correct BIS in which the new node is to be
inserted we first consider the situation in the suffix tree. We consider the following two cases.

\begin{itemize}
\item[(a)]
The new leaf $u$ in the suffix tree (representing the new suffix) is inserted as a child of an already
existing internal node $v$.
\item[(b)]
The new leaf $u$ in the suffix tree is inserted as a child of a new internal node $v$.
\end{itemize}

First, consider the case where the new leaf $u$ in the suffix tree is inserted as a child of an already
existing internal node $v$. If $v$ is either a $\sigma$-leaf or not a $\sigma$-node, then from
Observation~\ref{leaf} it is known that $leaf(v)$ and $u$ need to be in the
same BIS. So, we may "move" to the BIS in which $leaf(v)$ is positioned and traverse up from $leaf(v)$ to the root of the BIS (in
$O(\log |\Sigma|)$ time). If $v$ is a
branching-$\sigma$-node, then the BIS is found in $O(\log |\Sigma|)$ time
from $A_v$ (we show how to do this in Section~\ref{sub_sec:dynamic_branching_array}). Otherwise, $v$ is a $\sigma$-node which is not a $\sigma$-leaf and
not a branching-$\sigma$-node. In this case the correct BIS of
the two possible BISs can be found by examining the separating character
maintained in $v$.

Next, consider the case where the new leaf $u$ in the suffix tree is
inserted as a child of a new internal node $v$. Let $w$ be $v$'s
parent, and let $w'$ be $v$'s other child (not $u$). We consider this case in two steps. First we insert $v$ and then we insert $u$.
After we show how to update the suffix trist to include $v$, $u$ can be added as explained above.

We need to determine the status of $v$. Obviously, $v$ cannot be a
branching-$\sigma$-node. Moreover, the number of leaves in
$v$'s subtree is the same as the number of leaves in the subtree
rooted at $w'$ (as $u$ is currently being ignored). So, if $w'$ is
not a $\sigma$-node, $v$ is not a $\sigma$-node, and otherwise, $v$
is a non-branching-$\sigma$-node with a separating character that is the first
character of the label of edge $(v,w')$. The entire
process takes $O(\log |\Sigma|)$ time, as required.

\subsection{The Branching-$\sigma$-Node Array in a Dynamic Setting}\label{sub_sec:dynamic_branching_array}

The array $A_v$ for a branching-$\sigma$-node $v$ has to be adapted to work for the dynamic setting. First of all, it will be necessary to initialize the array when a node becomes a branching-$\sigma$-node (which naively takes $O(|\Sigma|)$ time). The other challenge is that the values of the array might be changing. Most notably, a suffix interval, represented by a BIS, might contain a node that changes to a $\sigma$-node thereby dividing the suffix interval. This requires updates to all the array locations of characters represented by the suffix interval. We will discuss these issues in more depth in Section~\ref{sub_sec:branching_sigma}. However, we present upfront the change in data structure format from that of the Suffix Tray. This will be useful for the discussion throughout Section~\ref{sec:status_change}.

The array $A_v$ in the (static) suffix tray contains a pointer for each character to its appropriate suffix interval. Specifically, pointers to the same suffix interval appear in all of the array cells that go to that suffix interval. This is fine in the static case but may be costly for text extensions since splitting a BIS can cause many pointers to need to be changed. Hence, we change our strategy and let an array cell representing character $a$ in the array point to the edge in the suffix tree leaving $v$ with $a$. During a traversal we implement the pointer to the appropriate BIS from a character $a$ by accessing the edge from $A_v$ at the location representing $a$. Say this edge leads to node $u$. Then from $leaf(u)$ we can access the appropriate BIS. Notice that while we will no longer need suffix interval nodes, we will still refer to them as if they exist while keeping in mind that accessing the appropriate BIS is done through the the appropriate edge (in a manner to be shown later) without the actual suffix interval node. The advantage is that, when a BIS representing a suffix interval is split, nothing in the array needs to be changed (only the new $\sigma$-node needs to be marked). We point out that the procedure we will describe to find the appropriate BIS will work regardless of the changing BISs.

We solve the problem of initializing $A_v$ as follows. Since we are in a dynamic environment, nodes will change status to become branching-$\sigma$-nodes as the text grows, thereby requiring new arrays. We discuss this later in Section~\ref{sub_sec:branching_sigma}. However, we do point out that it is helpful to maintain (throughout the algorithm) a balanced search tree at each node $v$ of the suffix tree, which we denote $BST_v$, containing the characters which exit the node (not all characters) with pointers to the edges exiting $v$ with the corresponding characters. We will use this data structure during the
time $A_v$ is being constructed. It will be shown later that this will not affect the query time.

\subsection{Search Outline}

We outline the procedure of the search for a pattern in a Suffix Trist. However, we point out that it is an outline and the reader is advised to reread the outline after reading Section~\ref{sec:status_change}.

\bigskip
\noindent
\underline{\bf Search}

\begin{enumerate}
\item{\bf $\sigma$-leaf $v$:} Go to the BIS that represents the suffixes in the subtree of $v$ in the suffix tree.
\item
{\bf Non-branching-$\sigma$-node $v$:} Compare $\tau$, the separating character, with $\rho$, the current character of the search.\\ If $\rho = \tau$ then go the child $v_i$ where $(v,v_i)$ is marked starting with $\tau$.\\
If $\rho < \tau$ then go to the left BIS and if $\rho > \tau$ then go to the right BIS.
\item
{\bf Branching-$\sigma$-node $v$:}
\begin{enumerate}
\item
If $A_v$ exists then go to $A_v(\rho)$ leading to edge $e=(v,u)$ exiting $v$ with $\rho$. Check the text on the edge for a match. If $u$ is a $\sigma$-node then continue the search in the suffix tree. If not, then go to $leaf(u)$, move to the BIS containing $leaf(u)$ and find the root of the BIS by going up the binary search tree of that BIS. Continue the search for the pattern within the BIS (as described in Section~\ref{sub_sec:BIS}).
\item
If $A_v$ is still under construction, compare $\rho$ with the first character on each of the edges leading to $v_i$ and $v_j$ (the first two children that became $\sigma$-nodes and caused $v$ to become a branching-$\sigma$-node). If $\rho$ matches one of them go to $v_i$ or $v_j$ accordingly, checking for a match on the way. If not, use $BST_v$ to find $\rho$ and then repeat the procedure described when $A_v$ exists.
\end{enumerate}

\end{enumerate}

\section{When a Node Changes Status}\label{sec:status_change}

At this stage of our description of the suffix trist we have assumed that we have an online suffix tree oracle. We have also described how to maintain the BISs that replace the suffix array intervals of the Suffix Tray. Finally we have shown how to find the correct BIS into which a new suffix is inserted during a text extension, and how to execute the insertion. We are left with the task of maintaining the auxiliary data on the nodes of the suffix tree, namely maintaining the data for $\sigma$-nodes, branching-$\sigma$-nodes and knowing when a $\sigma$-node is a $\sigma$-leaf. Of course, the status of a node may change since the text is now arriving online.

In~\ref{sub_sec:sigma_detection} we begin by describing how to detect when a node reaches the status of a $\sigma$-node. In~\ref{sub_sec:sigma_leaves} we describe what needs to be updated when a node becomes a $\sigma$-node. In~\ref{sub_sec:status_lose} we describe what happens when a $\sigma$-leaf "loses" its leaf status. In~\ref{sub_sec:branching_sigma} we describe the updates necessary when a $\sigma$-node becomes a branching-$\sigma$-node.

\subsection{Detecting a New $\sigma$-Node}\label{sub_sec:sigma_detection}

Let $u$ be a new $\sigma$-node and let $v$ be its parent. Just
before $u$ becomes a $\sigma$-node, (1) $v$ must have already been a
$\sigma$-node and (2) $u \in \{v_i,...,v_j\}$ and is associated with
an $(i,j)$-suffix interval represented by a suffix interval node $w$
that is a child of $v$ in the suffix trist. Hence, one will be able
to detect when a new $\sigma$-node is created by maintaining
counters for each of the (suffix tree) nodes $v_i,...,v_j$ to count
the number of leaves in their subtrees (in the suffix tree). These
counters are maintained on the suffix tree edges and each counter $v_k$ is indexed by the first
character on the edge $(v, v_k)$.

These counters only need to be maintained for nodes which are not $\sigma$-nodes and have a parent which is a $\sigma$-node. Thus, maintaining the counters can be done as follows. When a new leaf
is added into a given BIS of the suffix trist at suffix interval node
$w$ of the BIS, where $v$ is the parent of $w$, the counter of $v_k$ in the BIS needs to be increased, where $v_k$ is the one node (of the
nodes of $v_i,...,v_j$ of the suffix interval of the BIS) which is
an ancestor of the new leaf. The node $v_k$ can be found in
$O(\log |\Sigma|)$ time by using $BST_v$ with the search key being the character that appears at location $|label(v)|+1$ in the text; this character can be found in constant time with a direct access. Then, the counter of $v_k$ is incremented.

Notice that when a new internal node was inserted into the suffix
tree as described in Section~\ref{sec:suffix_trists}, it is possible that the
newly inserted internal node is now one of the nodes $v_i,...,
v_j$ for an $(i,j)$-suffix interval. In such a case, when the new
node is inserted, it copies the number of leaves in its subtree
from its only child (as was explained in Section~\ref{sec:suffix_trists} the newly inserted leaf is initially ignored and then treated as an independent insertion), as that child previously
maintained the number of leaves in its subtree. Furthermore, from
now on only the size of the subtree of the newly
inserted node is updated, and not the size of its child subtree.

Finally, when a node becomes a $\sigma$-node for the first time, the counters of all of its children will need to be updated. This is explained in further detail in Section~\ref{sub_sec:sigma_leaves}.

\subsection{Updating the New $\sigma$-Node}\label{sub_sec:sigma_leaves}

Let $u$ be the new $\sigma$-node (which is, of course, a
$\sigma$-leaf) and let $v$ be its parent. As discussed in the
previous subsection $u \in \{v_i,...,v_j\}$ where $v_i, ..., v_j$
are children of $v$ (in the suffix tree) and just before becoming a $\sigma$-node there
was a suffix interval node $w$ that was an $(i,j)$-suffix interval
with a BIS representing it.

Updating the new $\sigma$-node will require two things. First, the BIS is split into 3 parts; two new BISs and the new
$\sigma$-leaf that separates them. Second, for the new
$\sigma$-leaf the separating character is added (easy)
and a new set of counters for the children of $u$ is created (more
complicated).

The first goal will be to split the BIS that has just been updated
into three - the nodes corresponding to suffixes in $u$'s subtree,
the nodes corresponding to suffixes that are lexicographically
smaller than the suffixes in $u$'s subtree, and the nodes
corresponding to suffixes that are lexicographically larger than
the suffixes in $u$'s subtree.

As is well-known, for a given a value $x$, splitting a BST, balanced
search tree, into two BSTs at value $x$ can be implemented in
$O(h)$ time, where $h$ is the height of the BST (see Sections 4.1 and 4.2 in~\cite{Tar83}). The same is true
for BISs (although some straightforward technicalities are necessary to handle the
auxiliary information). Since the height of BISs is $O(\log
|\Sigma|)$ one can split a BIS into two BISs in $O(\log |\Sigma|)$
time and by finding the suffixes (nodes in the BIS) that correspond
to the rightmost and leftmost leaves of the subtree of $u$, one can
split the BIS into the three desired parts in $O(\log |\Sigma|)$
time. Fortunately, one can find the two nodes in the BIS in $O(\log
|\Sigma|)$ time by accessing the BIS directly through $leaf(u)$, and then walking up and down the BIS.

Recall that we need to maintain counters for nodes in the suffix tree which are not $\sigma$-nodes and have a parent which is a $\sigma$-node. So we need to initialize counters for all of the children of $u$ in the suffix tree. Denote these children of
$u$ by $u_1,...,u_k$. First notice that the number of suffixes in a
subtree of $u_i$ can be computed in $O(\log |\Sigma|)$ time by a
traversal in the BIS using classical methods of binary search trees.
It is now shown that there is enough time to initialize all the counters of
$u_1,...,u_k$ before one of them becomes a $\sigma$-node, while
still maintaining the $O(\log |\Sigma|)$ time bound per update.

Specifically, the counters will be updated during the first $k$
insertions into the BIS of $u$ (following the event of $u$
becoming a $\sigma$-node). At each insertion two of the
counters are updated. What is required is for the counters to be completely updated prior to the next time they will be used, i.e. in time to detect a new $\sigma$-node
occurring in the subtree of $u$. The following lemma is precisely
what is needed.

\begin{lemma}
Let $u$ be a node in the suffix tree, and let $u_1,...,u_k$ be $u$'s
children (in the suffix tree). Say $u$ has just become a
$\sigma$-node. Then at this time, the number of leaves in each of
the subtrees of $u$'s children is at most $|\Sigma|-k+1$.
\end{lemma}

\proof Assume by contradiction that this is not the case.
Specifically, assume that child $v_i$ has at least $|\Sigma|-k+2$
leaves in its subtree at this time. Clearly, the number of leaves in
each of the subtrees is at least one. So summing up the number of
leaves in all of the subtrees of $u_1,...,u_k$ is at least
$|\Sigma|-k+2+k-1=|\Sigma|+1$, contradicting the fact that $u$ just
became a $\sigma$-node (it should have already been a
$\sigma$-node).\qed

Since the size of the subtree of a child of $u$, say $u_i$, is no more than $|\Sigma|-k+1$ at least $k-1$ insertions will be required into the subtree of $u_i$, and hence into the subtree of $u$, before $u_i$ becomes a $\sigma$-node. Now, after each insertion into the subtree of $u$ (regardless of which child's subtree the new node was inserted into) we initialize the next two counters (by a continuous traversal on $BST_u$). Hence, after ${k \over 2}$ insertions all counters are initialized.

\subsection{When a $\sigma$-Leaf Loses its Status}\label{sub_sec:status_lose}

The situation where a $\sigma$-leaf becomes a non-leaf $\sigma$-node
is a case that has already been implicitly covered.
Let $v$ be a $\sigma$-leaf that is about to change its status to a
non-branching-$\sigma$-node which is not a leaf. This happens because one of its children
$v_k$ is about to become a $\sigma$-leaf. Notice that just before the
change $v$ is a suffix interval node. Hence,
the BIS representing the suffix interval needs to be split into
three parts, and the details are exactly the same as in
Section~\ref{sub_sec:sigma_leaves}. As before this is done in  $O(\log |\Sigma|)$ time.

\subsection{When a $\sigma$-Node Becomes a Branching-$\sigma$-Node}\label{sub_sec:branching_sigma}

Let $v$ be a $\sigma$-node that is changing its status to a
branching-$\sigma$-node. Just before it changes its status it had
exactly one child $v_j$ which was a $\sigma$-node. The change in
status must occur because another child (in the suffix tree), say
$v_i$, has become a $\sigma$-leaf (and now that $v$ has two children
that are $\sigma$-nodes it has become a branching-$\sigma$-node).

Assume, without loss of generality, that $v_i$ precedes $v_j$ in the
list of $v$'s children. Just before becoming a branching-$\sigma$-node, $v$ contained a
separating character $\tau$, the first character on the edge
$(v,v_j)$, and two suffix interval nodes $w$ and $x$, corresponding
to the left interval of $v$ and the right interval of $v$,
respectively. Now that $v_i$ became a $\sigma$-leaf $w$ was split
into three parts (as described in Section~\ref{sub_sec:sigma_leaves}).
So, in the suffix trist the children of $v$
are (1) a suffix interval node $w_L$, (2) a $\sigma$-leaf $v_i$, (3)
a suffix interval node $w_R$, (4) a $\sigma$-node $v_j$, and (5) a
suffix interval node $x$. Denote with $B_1, B_2$ and $B_3$
the BISs that represent the suffix interval nodes $w_L, w_R$ and
$x$.

The main problem here is that constructing the array $A_v$ takes too
much time, so one must use a different approach and spread the
construction over some time. A solution for which the $A_v$ construction charges its cost on future insertions (some of which may never
occur) is shown first. Then we show how to make it a worst-case solution. The charges will be over the insertions into the BISs $B_1, B_2$ and $B_3$ and will be described in the following lemma.

\begin{lemma}\label{lazybnode}
From the time that $v$ becomes a branching-$\sigma$-node, at least
$|\Sigma|$ insertions are required into $B_1, B_2$ or $B_3$ before
any node in the subtree of $v$ (in the suffix tree) that is not in
the subtrees of $v_i$ or $v_j$ becomes a branching-$\sigma$-node.
\end{lemma}

\proof Clearly, at this time, any node in the subtree of $v$ (in the
suffix tree) that is not in the subtrees of $v_i$ or $v_j$ has fewer
than $|\Sigma|$ leaves in its subtree. On the other hand, note that
any branching-$\sigma$-node must have at least $2|\Sigma|$ leaves in
its subtree, as it has at least two children that are
$\sigma$-nodes, each contributing at least $|\Sigma|$ leaves. Thus,
in order for a node in the subtree of $v$ (in the suffix tree) that
is not in the subtrees of $v_i$ or $v_j$  to become a
branching-$\sigma$-node, at least $|\Sigma|$ leaves need to be added
into its subtree, as required.\qed

This yields desired result, as one can always charge
the $A_v$ construction over its insertions into $B_1, B_2$ and
$B_3$. The crucial observation that follows from
Lemma~\ref{lazybnode} is that $A_v$ will be constructed
before a branching-$\sigma$-node that is a
descendant of $v$ but not of $v_i$ or $v_j$ is handled.

We now use a lazy approach that yields the worst-case result. We begin by using the folklore trick of initializing the array $A_v$
in constant time, see~\cite{aho74:_desig} (Ex. 2.12, page 71) and~\cite{Meh84} (Section III.8.1). Then every time an insertion takes place into one
of $B_1,B_2$ or $B_3$, one more element is added to the array $A_v$.
Lemma~\ref{lazybnode} assures that $A_v$ will be constructed
before a branching-$\sigma$-node that is a
descendant of $v$ but not of $v_i$ or $v_j$ is handled.

We note that in the interim, until $A_v$ is fully constructed, we use $BST_v$ during an indexing query as follows. First we compare the next character $\rho$ of the search with the first characters on $v_i$ and $v_j$.
If there is a match on either we continue the search in the direction of $v_i$ or $v_j$ accordingly. This is to make sure that the time spent at node $v$ is constant if we continue to $v_i$ or $v_j$. If this is not the case then we use $BST_v$ to search for the edge leaving $v$ starting with $\rho$. This can be done in $O(\log |\Sigma|)$ time. The following corollary of Lemma~\ref{lazybnode} assures us that we will use this access to a binary search tree only once in the whole search and, hence, the $O(\log |\Sigma|)$ time is additive.

\begin{cor}
Let $v$ be a branching-$\sigma$-node with children $v_i$ and $v_j$ being the first $\sigma$-nodes. Let $x$ be a branching-$\sigma$-node that is a descendant of $v$, but not of $v_i$ or $v_j$. Then $A_v$ is fully constructed.
\end{cor}

\proof By Lemma~\ref{lazybnode} there must be at least $|\Sigma|$ insertions into the subtree of $v$ that is not a subtree of $v_i$ or $v_j$ before $x$ can be a branching-$\sigma$-node. By then $A_v$ is fully constructed. \qed

We can finally conclude with the following theorem.

\begin{theorem}
Let $S$ be a string over an alphabet $\Sigma$. The suffix trist of
$S$ is (1) of size $O(n)$, (2) supports text extensions in time
$O(\log|\Sigma|)+ extension_{ST}(n,\Sigma))$ time (where
$f_{ST}(n,\Sigma)$ is the time for a text extension in the
suffix tree) and (3) supports indexing queries (of size $m$) in
$O(m+\log|\Sigma|)$ time.
\end{theorem}

\section{Conclusions}

We have shown how to create and maintain linear sized indexing structures that allow searches in $O(m+\log|\Sigma|)$ time for $m$ length patterns. We have shown this both for the static and online scenarios. In the static version we have produced the Suffix Tray data structure and in the online version we have produced the Suffix Trist data structure. The advantage of the Suffix Tray is its simplicity, while allowing fast searches. The Suffix Trist, although more involved, is a clean generalized dynamization of the Suffix Tray. One future challenge is in allowing deletions to take place in addition to the insertions. For example, one may be interested in a generalized suffix trist managing a collection of texts so you can insert new texts and delete texts from the collection.

\end{document}